# Dynamic Multi-Layer Signature Based Intrusion Detection System Using Mobile Agents

Mueen Uddin[1,] Kamran Khowaja[2] and Azizah Abdul Rehman[3]

Department of Information System, UTM, Malaysia
Mueenmalik9516@gmail.com , azizahar@utm.my

Faculty of Computer Science, Isra University Pakistan
kamran@isra.edu.pk

## ABSTRACT

*Intrusion detection systems have become a key component in ensuring the safety of systems and networks. As networks grow in size and speed continues to increase, it is crucial that efficient scalable techniques should be developed for IDS systems. Signature based detection is the most extensively used threat detection technique for Intrusion Detection Systems (IDS). One of the foremost challenges for signature-based IDS systems is how to keep up with large volume of incoming traffic when each packet needs to be compared with every signature in the database. When an IDS cannot keep up with the traffic flood, all it can do is to drop packets, therefore, may miss potential attacks. This paper proposes a new model called Dynamic Multi-Layer Signature based IDS using Mobile Agents, which can detect imminent threats with very high success rate by dynamically and automatically creating and using small and efficient multiple databases, and at the same time, provide mechanism to update these small signature databases at regular intervals using Mobile Agents.*

## KEYWORDS

*Intrusion Detection Systems, Snort, Mobile Agents, Aglets, Snort Rules*

## 1. INTRODUCTION

The current internet faces escalating threats form more sophisticated, intelligent and automated malicious codes. In the past, we have seen computer worms spread themselves without any human interaction and launched the most destructive attacks against computer networks. As an example, in January 2003, the SQL Slammer worm, also known as sapphire, was released into the internet exploiting a weakness into Microsoft SQL servers. In only 10 minutes the worm spread worldwide consuming massive amount of bandwidth and bringing down 5 of the 13 root DNS servers.

Amongst worms defensive mechanisms, Intrusion Detection systems (IDS) are the most widely deployed techniques that utilize the self-duplicating repetitive nature of computer worms to detect the patterns and signatures of theses malicious codes in the network traffic. These systems based on the parameters used for detection, can be broadly divided to signature based and anomaly based systems.

### 1.2 Signature-based IDS

Signature-based detection is normally used for detecting known attacks. There are different definitions of attack signatures. In this paper, the main discussion will focus on content signatures, which represent a string of characters that appear in the payload of attack packets. No knowledge of normal traffic is required but a signature database is needed for this type of detection systems. For worm detection, this type of system does not care how a worm finds the target, how it propagates itself or what transmission scheme it uses. The system takes a look at the payload and identify whether or not it contain a worm.

One big challenge of signature-based IDS is that every signature requires an entry in the database, and so a complete database might contain hundreds or even thousands of entries. Each packet is to be compared with all the entries in the database. This can be very resource- consuming and doing so will slow down the throughput and making the IDS vulnerable to DoS attacks. Some of the IDS evasion tools use this vulnerability and flood the signature signature-based IDS systems with too many packets to the point that the IDS cannot keep up with the traffic, thus making the IDS time out and drop packets and as a result, possibly miss attacks [1]. Further, this type of IDS is still vulnerable against unknown attacks as it relies on the signatures currently in the database to detect attacks.

## 1.2 Anomaly-based IDS

The signature of a new attack is not known before it is detected and carefully analyzed. So it is difficult to draw conclusions based on a small number of packets. In this case, anomaly-based systems detect abnormal behaviors and generate alarms based on the abnormal patterns in network traffic or application behaviors. Typical anomalous behaviors that may be captured include 1) misuse of network protocols such as overlapped IP fragments and running a standard protocol on a stealthy port; 2) uncharacteristic traffic patterns, such as more UDP packets compared to TCP ones, and 3) suspicious patterns in application payload.

The big challenges of anomaly based detection systems are defining what a normal network behavior is, deciding the threshold to trigger the alarm, and preventing false alarms. The users of the network are normally human, and people are hard to predict. If the normal model is not defined carefully, there will be lots of false alarms and the detection system will suffer from degraded performance.

## 2. PROPOSED WORK

From the above discussions and analysis, it becomes very much obvious that anomaly-based IDS have huge risk in generating high false positives, while in contrast, signature-based IDS are less susceptible to generate false alarms because the decision to generate an alarm is based on the signatures detected and does not require any knowledge of the normal traffic. However, signature-based IDS have their own limitations.

1. Every signature requires an entry in the database, and each packet needs to be compared with all the entries in the database. This may potentially slow down the throughput of the systems.

2. Signature-based IDS is vulnerable against newly emerging attacks.

For the second issue, there are already proposals and systems such as [6] that use a complementary payload-based anomaly detection system to detect new attacks, create signatures, and provide the new signatures to the signature-based IDS for the detection of the new threats.

Our focus in this paper will be addressing the first limitation of the signature-based IDS. In this paper, we proposed a new dynamic Multi-layer model for signature-based IDS with Mobile Agents. Our proposed model consists of multiple IDS systems deployed in different layers, and each is contained with small signature database. There is also a large complementary signature database containing all the entries of signatures detected during training period. We present a simple and automatic way to decide the set of rules and signatures to be deployed in the different IDS and continuously update the rules based on the usage pattern. Mobile Agents are used to perform the updating process of small signature databases. Using this model, we expect to optimize the detection rate of frequent threats to the network as well as providing means to detect uncommon attacks to the network.

## 3. RELATED WORK IN ADDRESSING LIMITATIONS OF SIGNATURE-BASED IDS

Signature based IDS systems require that their data bases need to be updated regularly at different time intervals so as to detect the imminent threads generated on the network. This process is a quite time consuming and requires a quick underlying system to update the database. Two layer signature based model was proposed to address signature based detections with unequal databases. But this model doesn't have any mechanism for adding, removing or updating signatures in the large signature based database [7]. If the signature database is not updated timely, then new threats will not be detected using this model.

P. Wheeler [8], in his thesis, divides all efforts to address this issue in the following categories:

1. Improving content matching algorithms- signature matching is one of the most computationally intensive tasks of IDS. Methods such as Landmark Segment Method (LSM) [9] and modified Aho-Corasick algorithm [10] have been proposed to reduce the execution time of signature matching.

2. Parallel processing – parallelism can be achieved at different levels: a) at node level by running a subset of rules on different nodes to keep the database size small; b) at component level by running different components of the IDS, such as preprocessors and rules engine, on different nodes to distribute the load; and c) at sub-component level by dividing a particular component of the IDS into n parallel instances, with each instance responsible for 1/n of the incoming packets.

While all the above efforts are geared toward improving the IDS throughput, they either require significant modifications to existing IDS by including new content matching algorithms, or require dedicated hardware resources to achieve acceleration, or need to coordinate tasks and aggregate output from parallel processing units. The proposed model is very simple, less resource demanding, yet can be used in combination with the above schemes. This model updated large complementary signature based database regularly in different time intervals using mobile agents. These mobile agents automatically add, remove and update large signature database without consuming time and resources. Hence proposed model doesn't add any additional overhead over the system performance and cost.

# 4. DYNAMIC MULTI-LAYER SIGNATURE-BASED IDS USING MOBILE AGENTS

The work proposed in this paper is motivated by the fact that it is easy and less time consuming to update small signature databases compare to large complementary signature database continuously from time to time. By doing this we can also improve the throughput of signature-based IDS, since a packet needs to be matched with less number of signatures in small signature database compare to one with huge number of signatures. This idea is not new, and in fact, it has been suggested by an installation note to system administrators of Snort [11]. To turn this idea into an effective one, we need to address three major issues.

1.  How to decide whether a given signature is likely to be helpful for possible attacks? We need systematic guidelines whether to remove a given signature. Currently, configuring the signature database is still a manual trial and error process of disabling some signatures and adding them back in after missing some attacks.

2.  What to do if we make the wrong choice and classify a useful signature as unlikely and remove it from the database? How to protect the network in this case?

3.  What to do once a new service or protocol is added to the network? We cannot completely rely on the administrators to remember to manually add the corresponding signatures to the database. This process is labor-intensive and can be error prone.

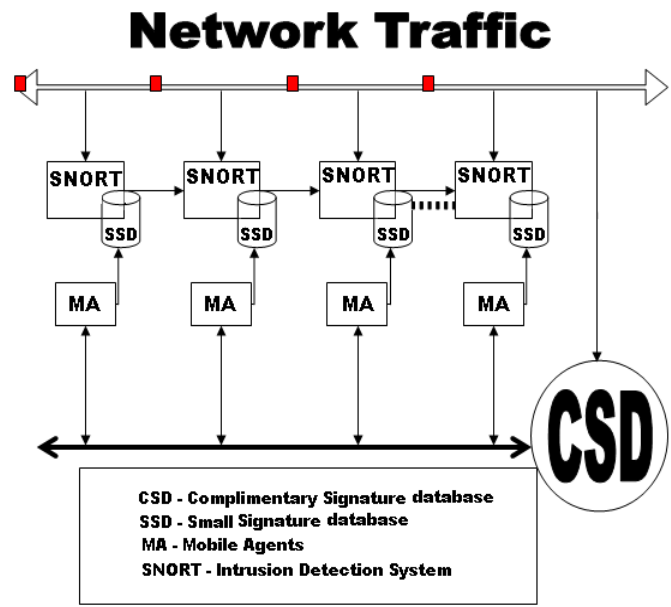

Figure1. Dynamic Multi layer signature based IDS using Mobile Agents

This paper addresses all of the above issues by proposing a new model based on small signature databases called Dynamic Multi Layer model. The proposed model consists of multiple smaller databases containing the most frequent attack signatures on the IDS, and a bigger complementary signature database containing thousands of signatures used to update smaller databases from time to time using mobile agents. By distributing the signature database between multiple nodes, it is important to consider two main aspects of the proposed model. 1) The signature database size on the nodes should not always be equal; it depends on the algorithm to update the small signature databases and also to keep the database size varying all the time. 2) The size of the databases on nodes dynamically changes to optimize the detection rate on more likely imminent threats.

The Implementation of proposed model comprises of two main phases:

**1. Training Period:** during the training period the system gathers information about the current threats to the network.

**2. Running Phase:** during this phase the signature databases are updated based on their usage patterns.

The duration of the training period depends on the network conditions and can vary from few hours to several days. The longer the training period, it is more likely to have a better baseline of the most frequent attacks on the network.

Once the training period is done, we look at the alert logs in the database. Depending on the criteria and parameters that can be defined by the user, we identify the most frequent attacks based on the 1) minimum number of occurrences of a signature, 2) the age of the alert in the database, 3) and the maximum number of signatures that we would like to keep in the all the IDS.

After identifying the most frequent attacks, we create a signature database for all IDS containing signatures for the most frequent attacks and create a complementary signature database containing the remaining signatures recognizable by the system. After the initial training period, Mobile Agents are used to update the small signature database if some new signature is updated in large complimentary database. This work is done periodically at certain time intervals defined in the implementation of mobile agents.

It is very much important to emphasize here that the proposed model will not miss attacks frequently as compare to the previous models proposed. This is because small signature databases are updated constantly with most updated and latest signatures using mobile agents. The complementary IDS help to detect less frequent attacks that may arise by comparing the signatures with packets travelling on the network.

## 5. PROTOTYPE IMPLEMENTATION

For prototype implementation, we chose Snort [11] as our signature-based IDS platform. We configured Snort's output plug-ins to log the alerts in the MySQL database for easy access and queries. Snort stores its signatures in rule files referenced in the Snort's configuration file.

Further, we developed a simple algorithm using mobile agents as an engine to create the most frequent signature databases for the multiple IDS as well as generating the complementary signature database for the secondary IDS. The same algorithm runs on all IDS systems for certain intervals to keep the primary signature database constantly updated. This can be done by removing signatures that are no longer occurring frequently, and also adding any signature detected as a frequent alert by the secondary IDS. The pseudo code for our program is presented in Figure 2 below. The algorithm accepts three input parameters: MinFreq specifying the minimum number of attack occurrences to be considered as frequent, ValidTime setting the time beyond which the attacks seen are considered as valid and threatening, and MaxNum representing the maximum number of the signatures acceptable in all IDS.

1.      N = 0   # number of current signatures
2.      Query the MySQL database to retrieve
        the set of signatures detected, S.
3.      **for** every signature f in S **do**
4.      Freq = number of occurrences of f
5.      LTime = last detection time of f
6.      **if** N <= MaxNum and Freq >= MinFreq and Ltime >= ValidTime **then**
7.      remove the signature from the secondary database
8.      add the signature in multiple IDS
9.      N = N+1
10.     **endif**
11.     **endfor**
12.     restart multiple and complimentary  IDS

Figure 2 Algorithm to generate and update signature databases

# 6. EXPERIMENT RESULTS

This section of the paper describes and discusses in detail the experimental setup made for performing experiments and then and analyze the results generated after performing different experiments at different time intervals with different parameters.

## 6.1 Experimental Setup

The experiments were performed choosing two different hardware platforms to simulate attacks and run IDS, one more powerful than the other. The objective was to simulate DoS attacks on IDS systems by running IDS on the slower machine and attacking tools on the faster machine. The aggregate computational and networking resources of attackers usually overwhelm the resources on the IDS machine. In this case, we would like to evaluate the effects of having multiple smaller   signature databases and how effectively it helps in improving the throughput and decreasing   packet loss rate. A small packet loss rate directly leads to small possibility to miss real attacks that might be hidden in false positive storms.

The detailed hardware and software configurations of systems used for performing experiments are as follows.

Attacking System:
        512MB memory / Windows XP
        CPU: AMD Geode NX running at 1.4GHz
        10/100Mbps NIC

IDS System:
        256MB memory / Windows XP
        CPU: Pentium III running at 500 MHz

10 Mbps NIC

Table 1 Signatures Detected during Training Period

| Signature Name | Number of Occurrences |
|---|---|
| (spp_frag2) TTL Limit Exceeded (reassemble) detection | 2 |
| (portscan) UDP Portscan | 3 |
| ICMP Destination Unreachable Communication | 3 |
| Administratively Prohibited | 3 |
| (spp_frag2) Teardrop attack | 4 |
| MISC gopher proxy | 4 |
| (portscan) TCP Portscan | 5 |
| WEB-MISC Compaq Insight directory traversal | 11 |
| DDOS tfn2k icmp possible communication | 14 |
| ICMP Large ICMP Packet | 14 |
| WEB-CGI search.cgi access | 17 |
| WEB-MISC http directory traversal | 18 |
| TELNET SGI telnetd format bug | 19 |
| (http_inspect) DOUBLE DECODING ATTACK | 20 |
| FINGER query | 21 |
| BACKDOOR Q access | 26 |
| BACKDOOR SIGNATURE - Q ICMP | 28 |
| DDOS mstream client to handler | 29 |
| BAD-TRAFFIC udp port 0 traffic | 35 |
| SNMP request udp | 40 |
| DOS arkiea backup | 48 |
| (http_inspect) WEBROOT DIRECTORY TRAVERSAL | 49 |
| BAD-TRAFFIC tcp port 0 traffic | 84 |
| (http_inspect) OVERSIZE REQUEST-URI DIRECTORY | 96 |
| NETBIOS RFParalyze Attempt | 105 |
| (spp_rpc_decode) Incomplete RPC segment | 129 |
| FTP command overflow attempt | 163 |
| WEB-CGI Allaire Pro Web Shell attempt | 981 |
| WEB-CGI Armada Style Master Index directory traversal | 998 |
| WEB-IIS index server file source code attempt | 1525 |
| BAD-TRAFFIC same SRC/DST | 1754 |
| ICMP Echo Reply | 3409 |
| BAD-TRAFFIC loopback traffic | 21886 |
| (snort_decoder): Invalid UDP header, length field | <86006 |

## 6.2 Attacking Tools

During the training period, following tools were used IDSwakeup [12], Stick [13], Sneeze [14], and Nikto [15], to trigger alerts by the IDS system and create a baseline of the most frequent attacks on our network. Here we briefly describe each tool.

IDSwakeup is designed to test the functionality of the IDS by generating some common attack signatures to trigger IDS alarms.

Stick is another tool fed with Snort configuration files to reverse engineer threats and create packets with signatures in the same way as those detected by Snort as attacks. Stick can be used to test the functionality of IDS as well as be deployed as a stress tester. It can be also used as an IDS evasion tool by generating a lot of traffic, and camouflaging the real attacks in a flood of false positives.

Sneeze is a Perl-based tool that is very similar to Stick in terms of functionalities. It distinguishes itself from Stick by the fact that it can accept Snort's rules at runtime and dynamically generate attack packets, whereas Stick needs to be configured with Snort's rules at compilation time.

Nikto focuses on web application attacks by scanning and testing web servers and their associated CGI scripts for thousands of potential vulnerabilities.

## 6.3 Analysis of Results

The results of the signatures detected during the training period are shown in Table 1.

For performing experiment, we set the value of MinFreq to 1, i.e., we considered any attack that appeared at least once as a frequent attack. In addition, we set ValidTime to be a negative number so that all the threats detected in the training period are to be included in the database. Our program scanned all the rule files of Snort and created a new rule file called "signature.rule" containing the most frequently signatures detected during the training period to be referenced by the snort.conf file as the only signature rule file. In addition, our program created complementary rule files taking out the most frequently used signatures for the secondary IDS system.

Snort was restarted on both systems pointing to the new signature files.

To test the performance of all IDS, we conducted our experiment in two different scenarios. In the first scenario, we manually enabled 3211 signatures and attacked the network using Sneeze [14].

In the second scenario, we let our program do its job by enabling only the most frequent signatures (in this case 33) as shows in Table 2 which shows the results of our tests in regards of the effects on packet drop rate (throughput).

Our test results clearly show the difference in the performance of the IDS using small signature database comparing to just enabling all signatures. In our environment with our specific configuration, by reducing the size of the database almost 97 times (3211/33), on average, we were able to decrease the percentage of the dropped packets by 6.77 times. This is a major improvement reducing the possibility of a real worm attack sneaking in the midst of dropped packets by the IDS system while ensuring all imminent threats can be detected by the IDS system.

Table 2: Performance Measurements (Test 1)

| Scenario 1 (All rules enabled) | | Scenario 2 (Most frequent signatures enabled) |
|---|---|---|
| **TEST 1** | | |
| Number of packets received | 187436 | 237503 |
| Number of packets analyzed | 173039 | 236700 |
| Number of packets dropped by Snort | 14397 | 803 |

Figure 3: Performance Measurement (Test 1)

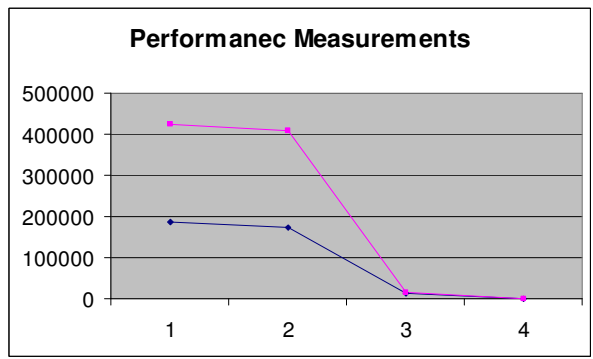

Table 3: Performance Measurements (Test 2)

| **TEST 2** | | |
|---|---|---|
| Packets received | 182688 | 193890 |
| Packets analyzed | 182688 | 190581 |
| Packets dropped by Snort | 15155 | 3309 |
| Percentage of packets dropped | 8.296% | 1.707% |

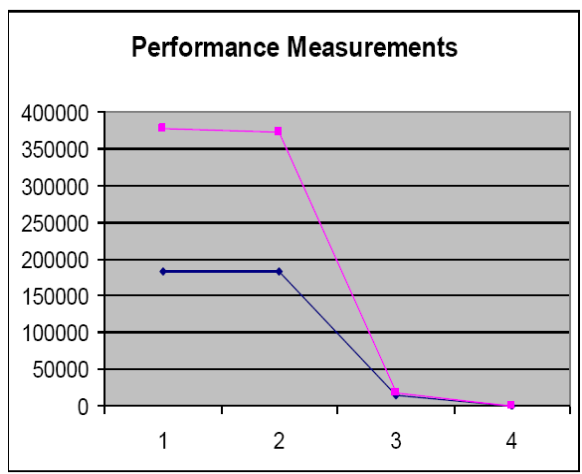

Figure 4: Performance Measurement (Test 2)

Table 4: Performance Measurements (Test 3)

| TEST 2 | | |
|---|---|---|
| Number of packets received | 182688 | 193890 |
| Number of packets analyzed | 182688 | 190581 |
| Number of packets dropped by Snort | 15155 | 3309 |
| Percentage of packets dropped | 8.296% | 1.707% |

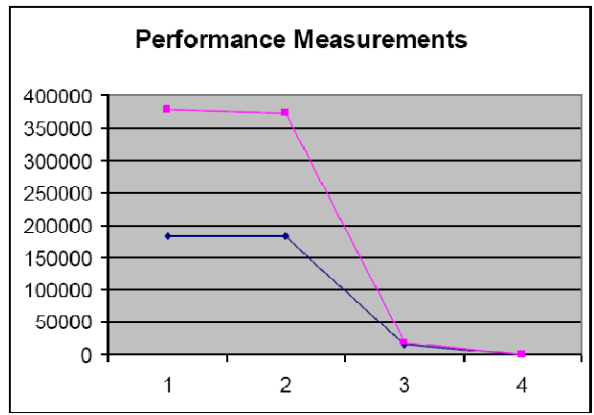

Figure 5: Performance Measurement (Test 2)

## 6.4 References

[1]     J. B. Raven Alder, Adam Doxtater, James Foster, Toby Kohlenberg, & Michael Rash, "Snort 2.1 Intrusion Detection," 2nd ed. Rockland, MA: Syngress (Distributed by O'Reilly and Associates), 2004.

[2]     C. W. C. Wong, D. Song, S. M. Bielski, & G. R. Ganger, "Dynamic Quarantine of Internet Worms," presented at 2004 International Conference on Dependable Systems and Networks, 2004.

[3]     C. S. D. Moore, G. M. Voelker, & S. Savage, Internet Quarantine: Requirements for Containing Self-Propagating Code," presented at IEEE INFOCOM, 2003.

[4]     S. C. Y. Tang, "Slowing Down Internet Worms," presented at The 24th IEEE International Conference on Distributed Computing Systems, 2004.

[5]     X. Q. D. Dagon, G. Gu, W. Lee, J. Grizzard, J.Levine, & H. Owen, "Honeystat: Local Worm Detection Using Honeypots," presented at Proceedings of the 7th Symposium on Recent Advances in Intrusion Detection, 2004.

[6]     D. F. Gong, "White Paper: Deciphering Detection Techniques: Part II Anomaly-based Intrusion Detection," Network Associates (McAfee Security), 2003.

[7].     M. Salour and Xiao Su, "Dynamic Two-Layer Signature-Based IDS with Unequal Databases", presented at proceeding of the International Conference on Information Technology (ITNG'07), 2007.

[8]     P. S. Wheeler, "Techniques for Improving the Performance of Signature-Based Network Intrusion Detection Systems," in Computer Science, Davis, CA: University of California, Davis, 2006.

[9]     T. C. Monther Aldwairi, and Paul Franzon, "Configurable String Matching Hardware for Speeding up Intrusion Detection," Special issue: Workshop on architectural support for security and anti-virus (WASSA), vol. 33, 2005.

[10]     B. C. S. Ryu, and J. Kim, "Design of Packet Detection System for High-Speed Network Environment," presented at Advanced Communication Technology, 2004. The 6th International Conference, 2004.

[11]     J. L. Sarang Dharmapurikar, "Fast and Scalable Pattern Matching for Content Filtering," presented at Proceedings of the 2005 symposium on Architecture for networking and communications systems ANCS '05, 2005.

[12]     Sourcefire,"Snort," http://www.snort.org, 2006.

[13]     S. Aubert, "IDSwakeup," http://www.hsc.fr/ressources/outils/idswakeup/index.html.en, 2000.

[14]     Coretez, "Stick," http://www.eurocompton.net/stick/projects8.html, 2001.

[15]     D. Bailey, "Sneeze," http://archives.neohapsis.com/archives/snort/2001-08/0180.html, 2001.

[16]     Sullo, "Nikto," http://www.cirt.net/code/nikto.shtml, 200.3

# 7.  CONCLUSION AND FUTURE WORK

In this paper, we introduced a new model of dynamic Multi-layer signature-based IDS using Mobile Agents to ensure that all the threats to the network are detected without compromising the performance. Our experiments proved a significant decrease in the packet drop rate, and as a result, a significant improvement in detecting threats to the network. Further, our proposed system can be improved by providing a more comprehensive and automated system that can distribute, add and remove the signatures across databases of multiple IDS systems based on the frequency of their appearance and their level of threat to the network. Finally, we believe more research needs to be done to determine the criteria to choose the optimal training period for a network.

**Authors**

**Mueen Uddin** is a PhD student at University Technology Malaysia UTM. His research interests include digital content protection and deep packet inspection, intrusion detection and prevention systems, analysis of MANET routing protocols, green IT, energy efficient data centers and Virtualization. Mueen has a BS & MS in Computer Science from Isra University Pakistan with specialty in Information networks.

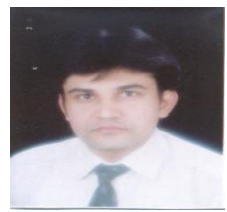

**Mr. Khowaja** has done MSc from Asian Institute of Technology in 2006 and MCS and BCS from Isra University in 2003 and 2001. He is teaching at university level for last 10 years and carrying out research in the field of Human Computer Interaction. He has supervised/co-supervised almost 15 Master degree projects and has published 8 research papers in various national and international conferences and journals.

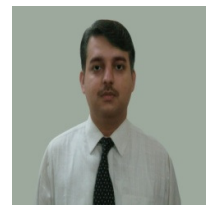

**Azizah Abdul Rehman** is an Associate Professor at University Technology Malaysia. His research interests include designing and implementing techniques for information systems in an organizational perspective, knowledge management, designing networking systems in reconfigurable hardware and software, and implementing security protocols needed for E-businesses. Azizah has BS and MS from USA, and PhD in information systems from University Technology Malaysia. She is 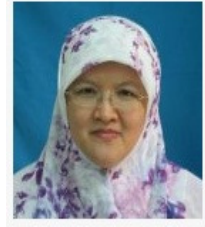 a member of the IEEE, AIS, ACM, and JIS. Azizah is a renowned researcher with over 40 publications in journals and conferences.